\documentclass[english,twocolumn]{revtex4-1}
\usepackage[T1]{fontenc}
\usepackage[latin9]{inputenc}
\usepackage{color}
\usepackage{amsmath}
\usepackage{amssymb}
\usepackage{graphicx}

\makeatletter
\usepackage{amsfonts}
\usepackage{babel}
\usepackage{babel}
\providecommand{\U}[1]{\protect\rule{.1in}{.1in}}

\usepackage{babel}

\makeatother

\usepackage{babel}
\begin{document}

\title{Theory for \emph{Inverse }\textit{Stochastic Resonance }\textit{\emph{in
Nature}}}

\author{Joaquín J. Torres$^{\dagger}$, Muhammet Uzuntarla$^{\ddagger}$
and J. Marro$^{\dagger}$\\
 {\small{}{}$\dagger$ Department of Electromagnetism and Physics
of the Matter, and }\\
 {\small{}{}Institute }\textit{\small{}Carlos I}{\small{}{} for
Theoretical and Computational Physics, }\\
 {\small{}{}University of Granada, 18071-Granada, Spain}\\
 {\small{}{}$\ddagger$ Department of Biomedical Engineering, Bulent
Ecevit University,}\\
 {\small{}{} Engineering Faculty, Zonguldak, Turkey }}
\begin{abstract}
The \textit{inverse stochastic resonance} (ISR) phenomenon consists
in an unexpected depression in the response of a system under external
noise, e.g., as observed in the behavior of the mean-firing rate in
some pacemaker neurons in the presence of moderate values of noise.
A possible requirement for such behavior is the existence of a bistable
regime in the behavior of these neurons. We here explore theoretically
the possible emergence of this behavior in a general bistable system,
and conclude on conditions the potential function which drives the
dynamics must accomplish. We show that such an intriguing, and apparently
widely observed, phenomenon ensues in the case of an asymmetric potential
function when the high activity minimum state of the system is metastable
with the largest basin of attraction and the low activity state is
the global minimum with a smaller basin of attraction. We discuss
on the relevance of such a picture to understand the ISR features
and to predict its general appearance in other natural systems that
share the requirements described here. Finally, we report another
intriguing non-standard stochastic resonance in our system, which
occurs in the absence of any weak signal input into the system and
whose emergence can be explained, with the ISR, within our theoretical
framework in this paper in terms of the shape of the potential function.
\end{abstract}
\maketitle

\section{Introduction}

Noise is ubiquitous in the real world, which has been a topic of significant
interest in the field of science and engineering applications. Much
effort have been devoted to understanding the source of noise and
the emergence mechanisms of noise induced phenomena as well as their
role in systems and devices. While noise was first considered to be
something that should be filtered out or reduced, it is now widely
accepted that noise can have a constructive role and enrich the stochastic
dynamics of nonlinear systems \cite{Pikovsky1997,LINDER2004,Faisal2008}.
A prominent example of this is stochastic resonance (SR) phenomenon
in which the signal to noise ratio in a nonlinear system under the
action of a weak input signal is maximized for a proper amount (not
too small nor too large) of noise \cite{Perc2007PhysRevE,Gammaitoni2009,McDonnell2009,Ozer2009,McDonnell2011}.
Under SR, a plot of the system response versus the ambient noise is
bell shaped, indicating that the correlation between the weak signal
and response is maximal around a moderate level of noise. A detailed
dynamical analysis has revealed that the existence of SR in natural
systems is due to the presence of kind of bistability on their driving
dynamics characterized, for instance, by the existence of a potential
function driving the dynamics with two minima separated by a finite
energy barrier. This barrier can be overpassed \textendash{} thus
inducing jumps of the activity among the two minima \textendash{}
thanks to a strong driving force or an efficiently high level of noise.
SR then occurs when, for very weak signals unable to jump, an optimal
amount of noise induces jumps that are thus correlated with the weak
signal \cite{Gammaitoni98}. 

The presence of such conditions of bistability during the activity
of actual neurons has been widely depicted \cite{lechner96,Paydarfar06,Engbers13}.
For instance,\textcolor{black}{{} the pacemaker activity of the squid
giant axon shows a pattern of switching on\textendash off activity
which depends on several features of a noisy input current and its
fluctuations \cite{Paydarfar06}. In this case, the authors illustrate
a kind of bistability in which noise induces different types of neuron
behavior, including repetitive firing, emergence of bursting and,
even more intriguing, the complete quietness of neural activity.}
Following these interesting findings, a series of theoretical studies
reported a new intriguing noise induced phenomenon in neural systems,
in which a minimum \textendash{} possibly zero \textendash{} occurs
in the average spiking activity of single neuron models for an optimal
amount of noise \cite{gutkininhibition2009,TuckwellPhysRevE09,Guo11,tuckwelleffects2011,Uzuntarlasolo13,Uzuntarla13,Yamakou2017}.
Such a noise induced behavior has also recently been found in neuronal
populations in biophysical realistic models with different network
coupling schemes \cite{Uzuntarla2017}. Following this, a \emph{double
inverse stochastic resonance} with two distinct minima has been reported
in the response of a Hodgkin-Huxley model neuron that receives synaptic
inputs subject to different types of short-term synaptic plasticity
\cite{Uzuntarla2017b}. Since the dependence of the neuronal system
response on the noise is the opposite to that in the SR mechanism,
by analogy, this phenomenon has been named ``inverse stochastic resonance''
(ISR). These previous studies have stated that co-existence of a stable
resting equilibrium and a stable spiking limit cycle during the model
neuron dynamics is the key factor for the emergence of interesting
noise induced effects. These theoretical findings have been complemented
with the first experimental evidence for ISR in an \emph{in vitro}
preperation of cerebral purkinje cells \cite{Buchin2016}. The authors
show in this work that ISR allows the Purkinje cells to operate in
different functional regimes depending on the variance of the neuronal
noise: the all-or-none toggle or the linear filter mode in cerebral
information processing. Furthermore, ISR might also play a critical
role in computational mechanisms that require reduced firing activity
without chemical inhibitory neuromodulation or, alternatively, when
other computational mechanisms require on-off bursts of activity.
On the other hand, apart from neural systems, a recent experimental
study reported the existence of ISR in nematic liquid crystals (NLC)
\cite{Huh2016}. This work shows that a proper arrangement of colored
noise intensity and its correlation time constant induces ISR in an
ac-driven electroconvection in NLC. 

These observations compelled us to develop a general theory, which
is presented in this paper, to explain the emergence of ISR in nature.
Our model considers the existence of two local minima separated by
an energy barrier for a potential function driving the activity of
a natural system, one corresponding to a low activity state, or Down
state, and the other corresponding to a high activity state, that
is the Up state. In addition, we assume that the low activity state
is the global minimum of the dynamics but it has a narrower basin
of attraction. On the other hand, the high activity state is a metastable
state but with a large basin of attraction. Note that these requirements
in our model impede in practice to observe ISR in pure bistable systems
with the two minima having their basins of attraction with the same
depth and size. It thus follows that ISR can only appear in natural
systems with metastable states.

\section{Model and Method}

\begin{figure}[tbh]
\centering{}\includegraphics[width=7cm]{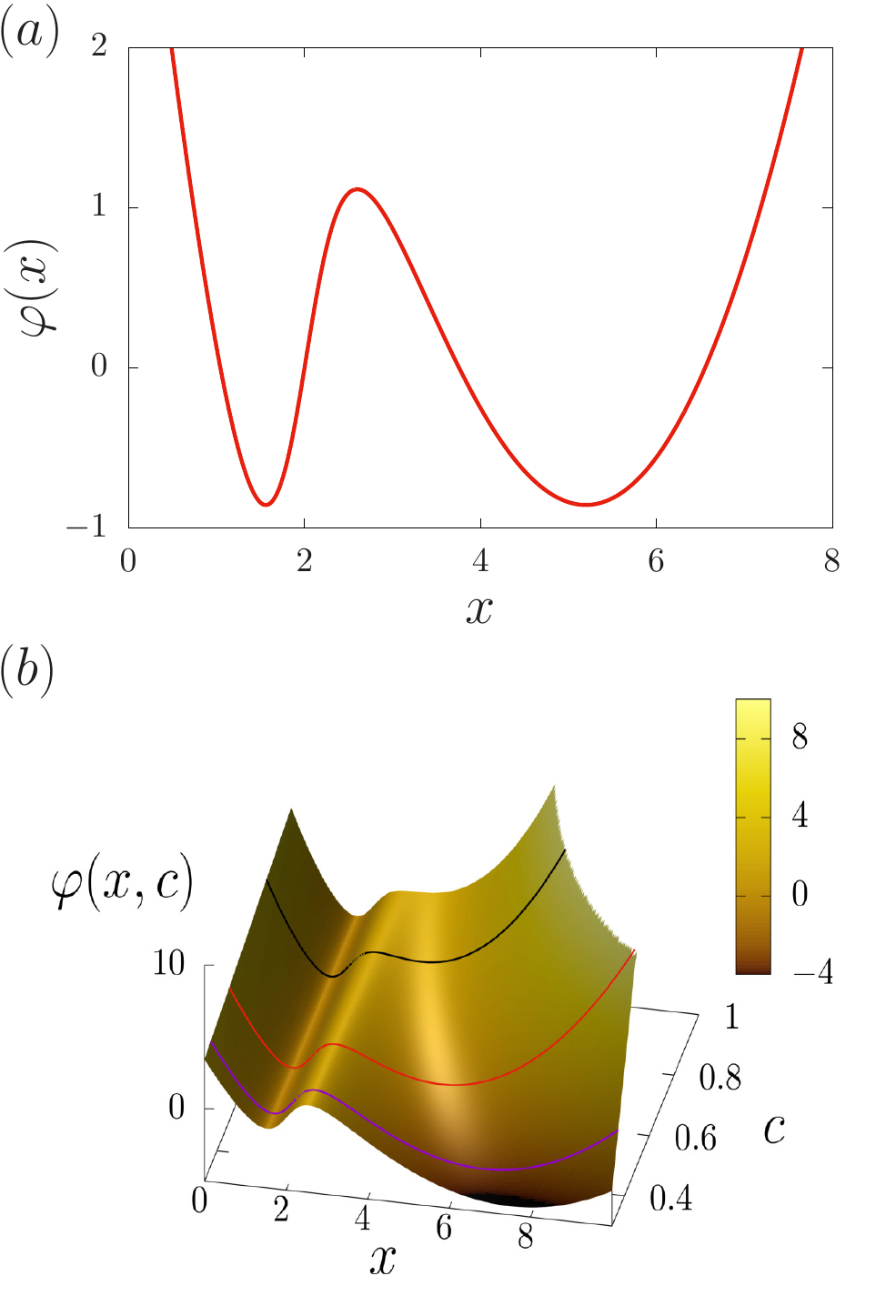}\caption{(a) Bistable potential $\varphi(x)$ with asymmetric basin of attractions
defined in equation (\ref{eq2}). Parameters are $a=3.367,\,b=2,\,c=0.5,\,$$d=-3.357,$
and $x_{0}=-2.$ (b) The same potential $\varphi(x,c)$ for $c\in(0.3,1)$
depicting transitions of the stability of the local minima as a function
of the parameter $c.$ Lines here correspond to $c=0.35$ (purple),
$c=0.5$ (red, the case illustrated in panel A corresponding to an
asymmetric bistable potential) and $c=0.8$ (black).}
\label{fig1} 
\end{figure}

Consider, for simplicity, a one dimensional dynamical system whose
activity or state is described by a variable $x(t)$ (representing
neural population activity, cell membrane voltage, chemical ion concentration,
etc) which follows the dynamics 
\begin{equation}
\frac{dx}{dt}=-\frac{\partial\varphi(x)}{\partial x}\label{eq:1}
\end{equation}
where the potential function is given, for instance, by 
\begin{equation}
\varphi(x)=a\arctan\left[b\left(x+x_{0}\right)\right]+c\left(x+x_{0}\right)^{2}+d\left(x+x_{0}\right),\label{eq2}
\end{equation}
which is a particular case of a class of familiar models for chemical
kinetics \cite{Becksei2528,TYSON2003221,asymmetricpot12}. This, which
yields 
\begin{equation}
\frac{dx}{dt}=-\frac{ab}{1+b^{2}\left(x+x_{0}\right)^{2}}-2c\left(x+x_{0}\right)-d,\label{eq3}
\end{equation}
produces, for the indicated set of parameter values, the shape asymmetry
of the local minima depicted in figure \ref{fig1}. 
This potential function depicts two locally stable minima separated
by a local maximum $x_{m}$, being $x_{1}$ the low activity one or
Down state, and $x_{2}$ being the high activity state or Up state,
in such a way that $x_{1}<x_{m}<x_{2}.$ We use this particular potential
since it is easy to tune between different shapes for it by changing
the value of a few parameters, as it is depicted in figure \ref{fig1}(b),
including the case of symmetric bistable potentials and asymmetric
ones. 

We are going to consider in the following the behavior of this system
under the action of a source of noise, i.e., the unidimensional Langevin
equation: 

\begin{equation}
\frac{dx}{dt}=-\frac{\partial\varphi(x)}{\partial x}+\eta(t).\label{eq:laneq-1}
\end{equation}
We assume here that the noise has the form of an additive Gaussian
term, $\eta(t)$ with zero mean, $\langle\eta(t)\rangle=0,$ and autocorrelation,
$\langle\eta(t)\eta(t^{\prime})\rangle=2D\delta(t-t^{\prime})$. 


\subsection{Limit of small noise}

Let us first consider the behavior of the system under the action
of a small noise. In addition, we assume a non-bistable potential
with two minima of different depth, one being the global minimum of
the dynamics and the other a metastable state. First, in the absence
of noise, after averaging over a number of trials with random initial
conditions $x_{0}\in(a,b),$ one has in the steady state that 
\begin{equation}
\langle x\rangle=x_{1}\,p\left(x_{0}<x_{m}\right)+x_{2}\,p\left(x_{0}>x_{m}\right).\label{eq:ave0}
\end{equation}
Here, $p(x_{0})=\left(b-a\right)^{-1}$ is the uniform distribution
of random initial conditions. It then follows that 
\begin{equation}
p\left(x_{0}<x_{m}\right)=\frac{x_{m}-a}{b-a},\quad p\left(x_{0}>x_{m}\right)=\frac{b-x_{m}}{b-a}.\label{eq:55}
\end{equation}
Consider next the system in the presence of a weak noise, in such
a way that, if its state at a given time is in the basin of attraction
of the metastable state (Up state), it has a probability of escape
during a time interval $dt$ equal to $p_{e}=dt/\left\langle t_{e}\right\rangle .$
The denominator here is the average escape time from that local minimum,
i.e., the inverse of the Kramer's escape rate probability, 
\begin{equation}
r_{K,2}=\frac{1}{2\pi}\left[\varphi^{\prime\prime}\left(x_{2}\right)\left\vert \varphi^{\prime\prime}\left(x_{m}\right)\right\vert \right]^{1/2}\exp\left(-\Delta\varphi_{2}/D\right),\label{eq:r2}
\end{equation}
from the high activity $x_{2}$ to the low activity $x_{1}$ values
of the minimum. $\Delta\varphi_{2}\equiv\varphi\left(x_{m}\right)-\varphi\left(x_{2}\right)$
is the potential barrier the state of the system has to overpass (see
for instance \cite{Riskenbook}) to escape from the metastable state.
Similarly, if the state of the system is around the low value minimum
$x_{1}$ it can jump to the high-value minimum $x_{2}$ with rate
probability 
\begin{equation}
r_{K,1}=\frac{1}{2\pi}\left[\varphi^{\prime\prime}\left(x_{1}\right)\left\vert \varphi^{\prime\prime}\left(x_{m}\right)\right\vert \right]^{1/2}\exp\left(-\Delta\varphi_{1}/D\right),\label{eq:r1}
\end{equation}
where $\Delta\varphi_{1}\equiv\varphi\left(x_{m}\right)-\varphi\left(x_{1}\right)$
is the potential barrier that the state of the system has to overpass
now starting around the minimum $x_{1}.$ The expressions \eqref{eq:r2}
and \eqref{eq:r1} are only valid when noise fluctuations are small
so that $\Delta\varphi\gg D$ \cite{Riskenbook}. 

Consider now that, at given time $t,$ the state of the system has
a probability $p(x_{t}<x_{m})\equiv\alpha_{t}$ to be around the minimum
at $x_{1},$ and a probability $p(x_{t}>x_{m})\equiv\beta_{t}$ to
be around the minimum at $x_{2},$ with $\alpha_{t}+\beta_{t}=1\forall t.$
Then, it follows that $\alpha_{0}=p(x_{0}<x_{m})$ and $\beta_{0}=p(x_{0}>x_{m}).$
One may now assume that both $\alpha_{t}$ and $\beta_{t}$ evolve
in time according to the equations $\frac{d\alpha_{t}}{dt}=r_{K,2}\beta_{t}-r_{K,1}\alpha_{t}$
and $\frac{d\beta_{t}}{dt}=r_{K,1}\alpha_{t}-r_{K,2}\beta_{t}$ due
to possible escapes from one minimum to the other and vice versa.
Then, one can easily solve the previous system of equations to obtain
\begin{equation}
\begin{array}{l}
\alpha_{t}=[r_{K,2}-(r_{K,2}-\lambda\alpha_{0})e^{-\lambda t}]/\lambda\\
\beta_{t}=[r_{K,1}-(r_{K,1}-\lambda\beta_{0})e^{-\lambda t}]/\lambda
\end{array}\label{eq:avet}
\end{equation}
where $\lambda=r_{K,1}+r_{K,2}$. We may compute the mean state or
activity of the system at each time $t$, which is approximately 
\begin{equation}
\langle x\rangle_{t}\approx x_{1}\alpha_{t}+x_{2}\beta_{t}.\label{eq:xt}
\end{equation}
At $t=0,$ one has that $\langle x\rangle_{0}\approx x_{1}\alpha_{0}+x_{2}\beta_{0},$
which coincides with (\ref{eq:ave0}). On the other hand, the steady
state of (\ref{eq:avet}) is $\alpha_{\infty}=\frac{r_{K,2}}{r_{K,1}+r_{K,2}}=p(x_{\infty}<x_{m}),$
$\beta_{\infty}=\frac{r_{K,1}}{r_{K,1}+r_{K,2}}=p(x_{\infty}>x_{m})$
which are the probability of the system to be in a state around the
minimum $x_{1}$ and $x_{2}$ respectively in the steady state.

\subsection{The limit of high noise }

Alternatively, in the presence of a noise high enough to overpass
the potential barrier, we can perform a different derivation to compute
the mean activity of the system in the steady sate. Starting with
the Langevin description \eqref{eq:laneq-1}, $x(t)$ is a random
variable with probability distribution $P(x,t)$ that obeys the Fokker-Planck
equation \cite{vankampen}
\begin{equation}
\frac{dP(x,t)}{dt}=-\frac{\partial}{\partial x}A(x)P(x,t)+D\frac{\partial^{2}}{\partial x^{2}}P(x,t).\label{eq:102}
\end{equation}
with $A(x)=-\frac{\partial\varphi(x)}{\partial x}.$ The steady-state
solution $P_{\infty}(x)=\lim_{t\rightarrow\infty}P(x,t)$ to this
equilibrium is known to be \cite{vankampen} 
\begin{equation}
P_{\infty}(x)=Ne^{-\varphi(x)/D},\label{eq:112}
\end{equation}
where $N=[\int e^{-\varphi(x)/D}dx]^{-1}$ is a normalization factor.
We can then compute the average of $\langle x\rangle$ using such
probability distribution and write it as a function of the potential
parameters and the noise level $D,$ that is, 
\begin{equation}
\langle x\rangle=\int_{\mathbb{R}}x\,P_{\infty}(x)=\frac{\int_{\mathbb{R}}x\,e^{-\varphi(x)/D}dx}{\int_{\mathbb{R}}e^{-\varphi(x)/D}dx}.\label{eq:equilibrium}
\end{equation}

\section{Results}

\subsection{Inverse stochastic resonance}

\begin{figure}[tbh]
\begin{centering}
\includegraphics[width=7cm]{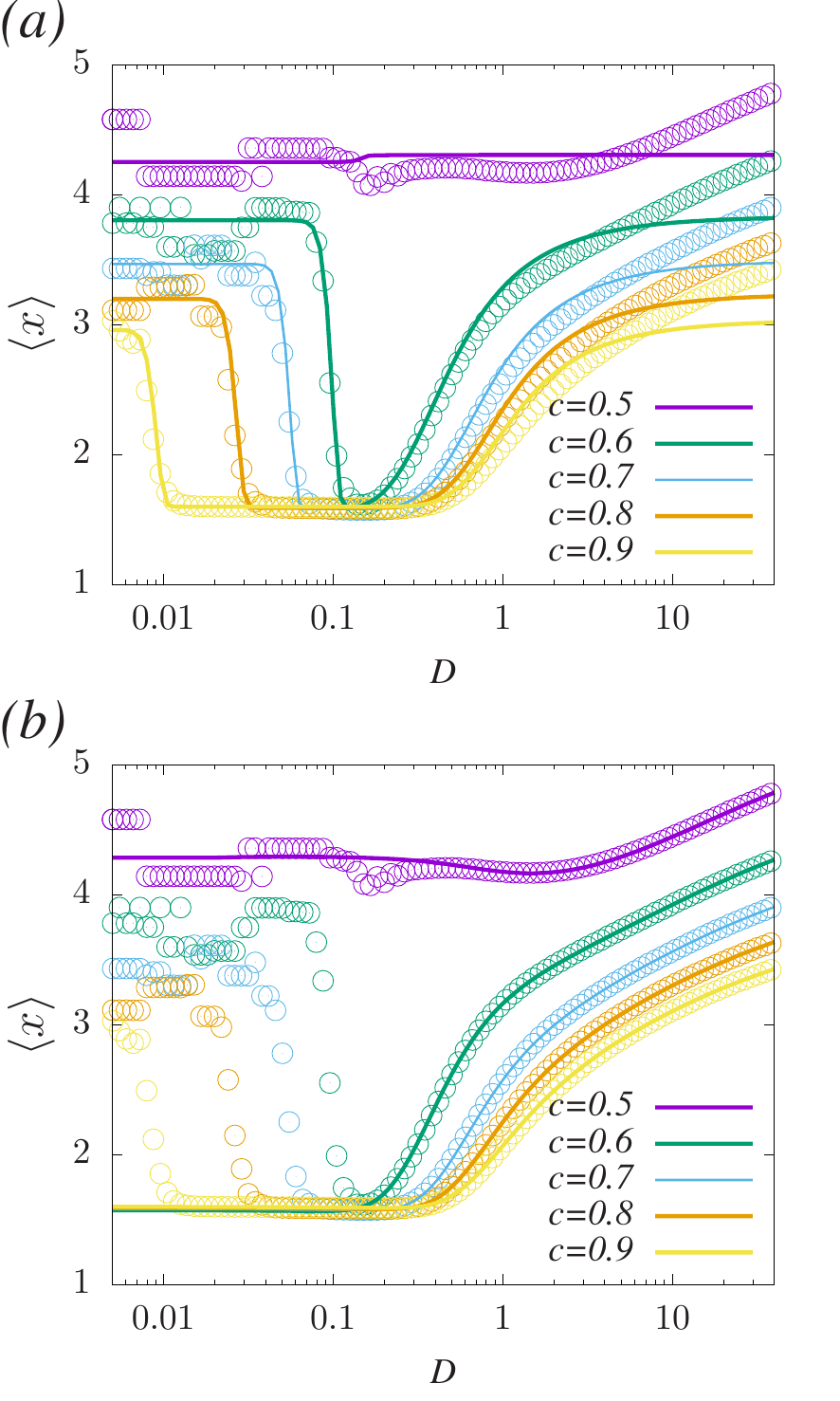}
\par\end{centering}
\centering{}\caption{(a) ISR curves obtained in simulations (open circles) compared with
the theory for low noise based in our Kramer's type theory (solid
lines), for different values of the potential function parameter $c$.
(b) ISR curves obtained in simulations (open circles) compared with
the equilibrium theory for high noise (solid lines), also for different
values of the potential function parameter $c$. Other model parameters
as in figure \ref{fig1}.}
\label{fig2} 
\end{figure}

Within the above model one may easily understand the shape and features
of the ISR curves previously reported in the literature. Figure \ref{fig2}
shows the level of agreement with the ISR curves computed in simulations
of the Langevin equation \eqref{eq:laneq-1} for different values
of the potential parameter $c$ (open circles), compared with the
value of $\langle x\rangle$ computed with the low-noise Kramer's
approximation from equation (\ref{fig2}) and depicted in panel (a)
(solid lines). The high-noise equilibrium theory using expression
(\ref{eq:equilibrium}) is shown in panel (b) (solid lines) also compared
with simulations of the Langevin equation (open circles). 
This shows that the Kramer's escape theory reproduces very well the
behavior of the ISR curve for low noise while fails for large noise.
The agreement for very low noise is due to the fact that in this case
the noise fluctuations cannot overpass the potential barrier in finite
time, so that the observed mean activity at zero noise is just the
result of an initial condition effect, and it can be computed through
the probabilities that the state of the system initially falls in
any of the basin of attractions of each minimum. That is, the zero
noise mean activity can be computed through the expression (\ref{eq:ave0}).
Note that, in this case the asymmetry between minima in the potential
function is what induces the zero noise level of the ISR curve to
be higher or lower. In fact, if the basin of attraction of the low
activity minimum is too narrow, then $p\left(x_{0}<x_{m}\right)\ll p\left(x_{0}>x_{m}\right)$
and therefore $\langle x\rangle\approx x_{2}.$ On the other hand,
if the high-activity minimum has a very narrow basin of attraction,
then $p\left(x_{0}<x_{m}\right)\ll p\left(x_{0}>x_{m}\right)$ and
therefore $\langle x\rangle\approx x_{1},$ which will impede the
appearance of ISR, since there is not a raising of the ISR curve toward
a higher value for very low noise.

When the noise increases but still remaining small, then there is
a non-zero probability, according with the Kramer's theory, that the
state of the system being initially in the basin of attraction of
the high-activity minimum could overpass the potential barrier due
to noise fluctuations and fall into the basin of attraction of the
low activity minimum. In this case, to reproduce the lowering of the
level of $\langle x\rangle$ for such noise levels, it is also important
that the low-activity minimum is deeper than the high-activity one,
in such a way that, for the same level of noise fluctuations, is easy
for the state of the system to overpass the potential barrier from
the high to the low activity minimum and harder the opposite. In such
conditions, system states being initially around $x_{2}$ could be
trapped around $x_{1}$ and, therefore, the $\langle x\rangle$ decreases
as simulations of ISR depict. Note that for this small non-zero level
of noise fluctuations, the behavior of $\langle x\rangle$ is mainly
determined by the rate probabilities for escaping from the potential
minima, and it can be computed through expression \eqref{eq:xt}.
In this case, the asymmetry of the potential with reference to the
depth of the minima is what determines the behavior of $\langle x\rangle$
as a function of the fluctuations intensity. In fact, such asymmetry
implies that $r_{K,2}\gg r_{K,1}$ and therefore $\alpha_{t}\gg\beta_{t},$
which makes $\langle x\rangle$ to reach a minimum value, that is,
the minimum of the ISR curve. 

For increasing values of the noise level, the expression \eqref{eq:xt}
can approximately explain the rising of $\langle x\rangle$ from the
minimum of the ISR curves since the probability to be trapped in the
low-activity minimum decreases. This occurs due to the fact that escapes
from such low-activity minimum to the high-activity one can also occur,
so both $r_{K,1}$and $r_{K,2}$ increase and therefore $\beta_{t}$
also increases and become comparable with $\alpha_{t}.$ However,
for large values of noise such tendency fails. This is because the
non-validity of the Kramer's theory which is based on the assumption
of non-negligible small current probability through the potential
barrier between the two minima, and for large noise such current probability
is negligible due to large size stochastic jumps ($\Delta\varphi\ll D$).
In such a case, we can consider that the system quickly reaches an
equilibrium condition where the state of the system is randomly exploring
its entire phase space. We then need to compute the steady-state probability
$P_{\infty}(x)$ to evaluate $\langle x\rangle$ using the expression
\eqref{eq:equilibrium}. Figure 2(b) clearly illustrates that such
equilibrium theory can exactly reproduce the shape of the ISR curves
for large noise values and any value of the potential parameter $c$.
However, the equilibrium theory incorrectly predicts, for low noise
values the lower value for the mean activity of the system $\langle x\rangle\approx x_{1}.$
The reason is that such minimum $x_{1}$ corresponds to the global
minimum of the dynamics and in steady-state conditions the system
is in equilibrium around such minimum. However, the equilibrium theory
does not account for the possibility that, for small noise and for
some initial conditions, the state of the system can be trapped during
a long time (that increases as the noise fluctuations decrease) around
the metastable high-activity minimum, which is the responsible for
the rising of $\langle x\rangle$ for small noise in the ISR curve
in simulations. 

\begin{figure}
\begin{centering}
\includegraphics[width=7cm]{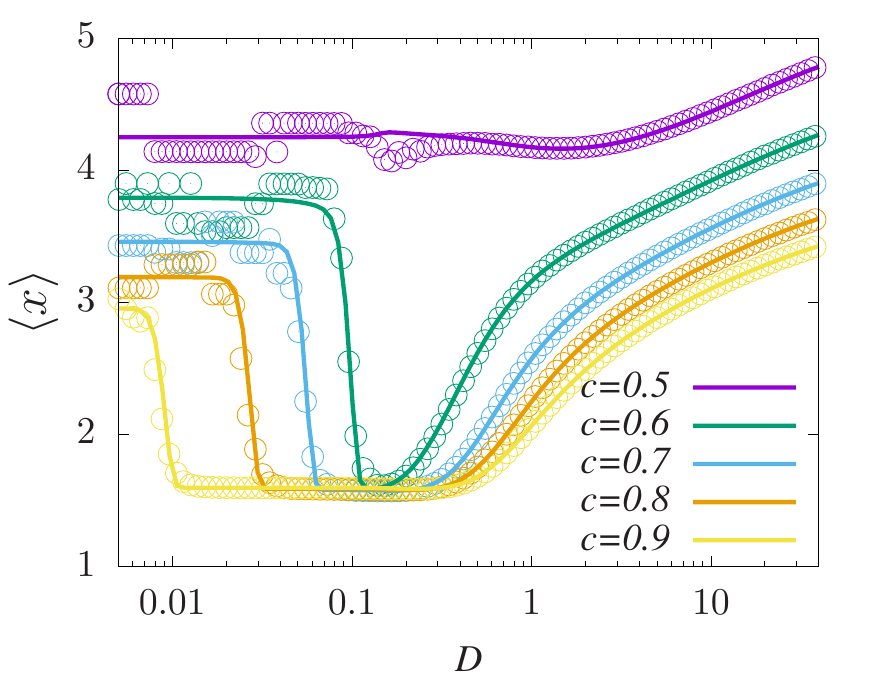}
\par\end{centering}
\caption{Emerge of ISR in an asymmetric non bistable potential. Lines corresponds
to the theory in this paper and open circles correspond to simulations
of the Langevin equation (\ref{eq:laneq-1}) for several values of
the parameter $c$ controlling the deep and asymmetry of the potential
function $\varphi(x).$ In all cases $D_{0}=0.15.$ Other model parameters
as in figure \ref{fig1}.}

\label{completa}
\end{figure}

One can account for both theoretical results \textendash{} that is
the small noise and high noise previous theories \textendash{} assuming
that there is a validity crossover between both theories. Therefore,
one can assume for all levels of noise that the mean activity of the
the system is given by
\begin{equation}
\langle x\rangle=\langle x\rangle_{L}[1-\xi(D)]+\xi(D)\langle x\rangle_{H}\label{eq:final}
\end{equation}
where the labels L and H indicate, respectively, the theories for
low and high noise, and $\xi(D)=\frac{1}{2}+\frac{1}{2}\mbox{\ensuremath{\tanh}\ensuremath{[10(D-\ensuremath{D_{0}})].}}$
This is function of noise such that, for $D>D_{0},$ $\xi(D)\approx1$
indicating that the high noise theory becomes the important one while
for $D<D_{0},$ $\xi(D)\approx0$ so that the low-noise theory is
the important one. Figure \ref{completa} depicts the behavior of
$\langle x\rangle$ as a function of noise level $D$ as given by
the expression (\ref{eq:final}). This shows good agreement between
the full theory (solid lines) and simulations of the Langevin equation
(\ref{eq:laneq-1}) for different values of the potential parameter
$c.$ The figure also clearly depicts that only for $c>0.5$ ISR emerges.
This corresponds to situations in which the system is not bistable
and the low activity minimum becomes the global minimum of the dynamics
with the high activity one becoming a metastable state. For $c\le0.5,$
the ISR behavior is lost since the global minimum is the high activity
one while the low activity minimum becomes a metastable state in such
a way that, for any value of the noise, the probability of the state
of system to be around such minimum is very low and, therefore, $\langle x\rangle$
becomes large for all values of $D.$

\subsection{Non-standard stochastic resonance or noise induced activity amplification}

\begin{figure}
\begin{centering}
\includegraphics[width=7cm]{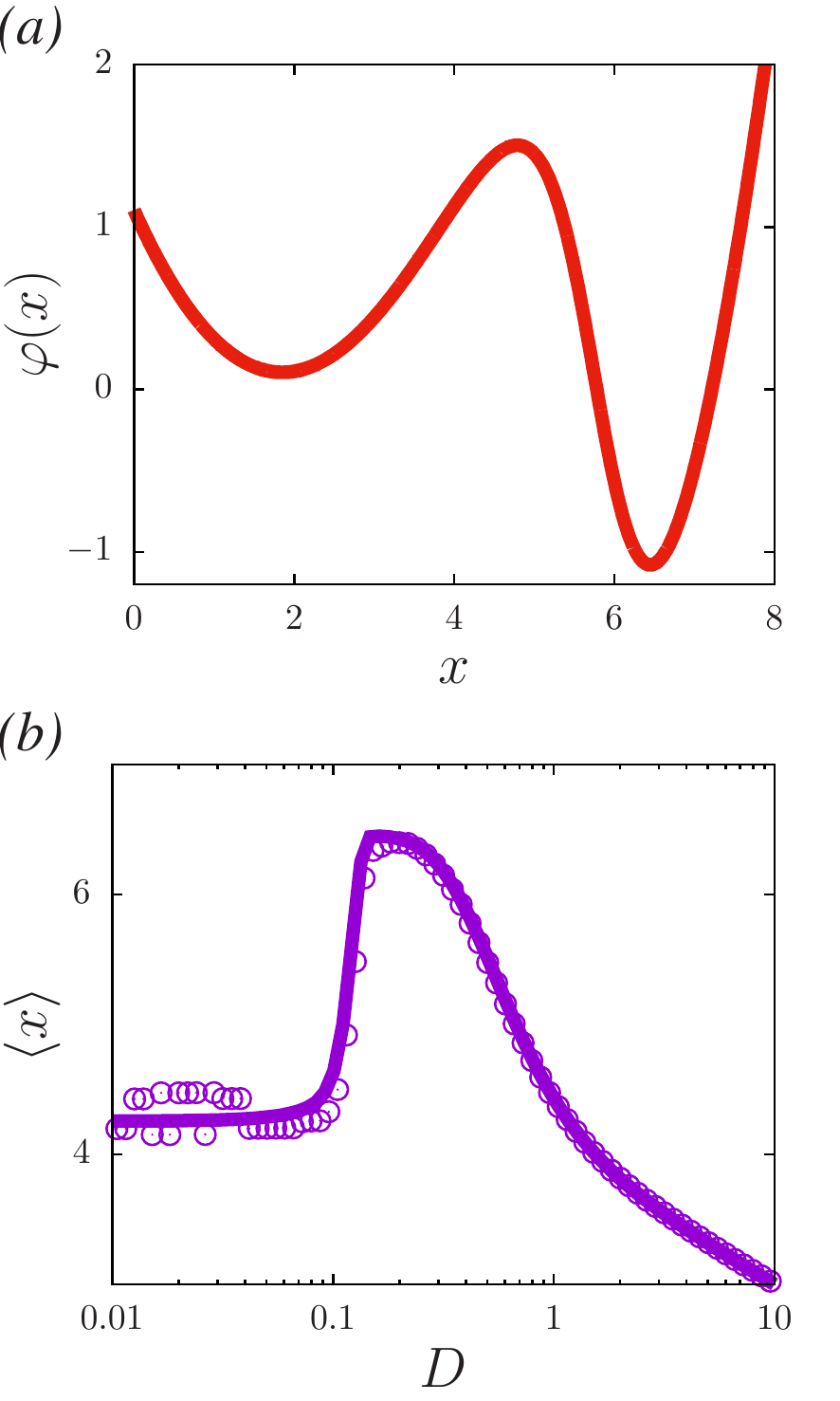}
\par\end{centering}
\caption{Emergence of noise induced activity amplification or non-standard
stochastic resonance. (a) This shows the asymmetric potential $\varphi(x)$
with two minima with different depth and size of the basin of attraction
for each minimum. Here, the low activity minimum is a metastable sate
with a large basin of attraction, and the high activity one is the
global minimun of the dynamics and has a small basin of attraction.
Potential function parameters are $a=4.367,\,b=-1.28,\,c=0.327,\,d=2.79$
and $x_{0}=-5.78.$ (b) Behavior of the $\langle x\rangle$ as a function
of the noise parameter $D,$ for the case of the potential function
illustrated in panel A. The graph depicts an amplification of the
observed mean activity for intermediate values of the noise parameter,
which resembles stochastic resonance. In this case $D_{0}=0.15.$
The solid line corresponds to our theory and data points (open circles)
to simulations of the Langevin equation \eqref{eq:laneq-1}.}

\label{fig4}
\end{figure}

Our theory also predicts a non-standard stochastic resonance (NSSR)
in our system , that is, the mean activity $\langle x\rangle$ may
present a maximum as a function of noise level even in the absence
of a weak signal in the right-hand side of equation \eqref{eq:laneq-1}.
The only requirement for this is to have dynamics \eqref{eq:laneq-1},
as before, with a potential function with two minima, but now with
the low-activity one being the metastable state and with the larger
basin of attraction and the high activity one being the global minimum
of the dynamics with a narrower basin of attraction, as depicted in
figure \ref{fig4}(a). The resulting behavior of $\langle x\rangle$
as a function of noise parameter $D$ is depicted in figure \ref{fig4}(b),
where a maximum in the mean activity is observed at intermediate values
of the noise parameter. This is quite similar to the typical stochastic
resonance curves widely reported in the literature. Note, however,
that the mechanism responsible for this behavior differs from that
of the typical SR behavior, where there is enhancement of the correlation
between the activity of a nonlinear system and a weak external input
(see for instance \cite{Gammaitoni98}). 

\section{Final discussion}

In this paper, we present a simple and general theory to explain the
emergence of ISR in a variety of natural situations. Our theory predicts
that ISR will occur in any natural system whose dynamics can be interpreted
as following from some potential function with two minima, one of
them being metastable (the one corresponding to the higher activity)
and the other being the global minimum (i.e., corresponding to the
lowest activity). In addition, a requirement for the appearance of
ISR is that the high activity minimum must have the larger basin of
attraction while the low activity one must be the global minimum of
the dynamics. We predict that any natural system that meets these
requirements is a potential candidate in which ISR should perhaps
be observed.

The theory presented here also predicts the existence of a non-standard
stochastic resonance phenomenon if the global minimum of the dynamics
corresponds to the higher activity one whereas the local or metastable
minimum corresponds to the lower activity one. The resulting stochastic
resonance is a non-standard one in the sense that the dynamics does
not need a weak signal input to induce a maximum in the average activity
of the system. In fact the observed maximum in the average activity
does not indicate here a correlation between system input and output
response. Thus, it can be considered as an activity amplification
phenomenon induced by a proper arrangement of the noise. We predict
that this type of non standard SR can emerge in any biological or
other natural system that meets the basic requirements as explained
above. 

Summing up, due to the simplicity and the generality of the theory
presented here, our study, firstly, can be easily extended to include
other types of noise, e.g. colored noise, to analyze their influence
on ISR features and, secondly, it can be useful to further explore
the possible implications that both ISR or non-standard SR could have
in the behavior of many different natural and physical systems. 

\section{Acknowledgments}

We acknowledge the Spanish Ministry for Science and the ``Agencia
Española de Investigación'' (AEI) for financial support under grant
FIS2017-84256-P (FEDER funds). We also thank fruitful comments from
P. L. Garrido.

\end{document}